\documentclass[hyper,letterpaper,12pt]{article}
\usepackage{a4wide}
\usepackage{amsmath}
\usepackage[
      colorlinks=true,
      linkcolor=blue,
      urlcolor=blue,
      filecolor=black,
      citecolor=blue,
            ]{hyperref}
\usepackage{color}
\usepackage{graphicx}
\usepackage{amsfonts}
\usepackage{amssymb}
\usepackage{epstopdf}

\def\beq{\begin{equation}}
\def\eeq{\end{equation}}

\begin{document}
\title{\bf \Large Competition between s-wave order and d-wave order in holographic superconductors}

\author{\large
~~Li-Fang Li$^1$\footnote{E-mail: lilf@itp.ac.cn}~,
~Rong-Gen Cai$^{2,3}$\footnote{E-mail: cairg@itp.ac.cn}~,
~~Li Li$^2$\footnote{E-mail: liliphy@itp.ac.cn}~,
~~Yong-Qiang Wang$^4$\footnote{E-mail: yqwang@lzu.edu.cn}
\date{\today}
\\
\\
\small $^1$State Key Laboratory of Space Weather, \\
\small Center for Space Science and Applied Research, Chinese Academy of Sciences,\\
\small Beijing 100190, China.\\
\small $^2$State Key Laboratory of Theoretical Physics,\\
\small Institute of Theoretical Physics, Chinese Academy of Sciences,\\
\small Beijing 100190,  China.\\
\small $^3$ King Abdulaziz University, Jeddah, Saudi Arabia.\\
\small $^4$Institute of Theoretical Physics, Lanzhou University, Lanzhou, 730000, China.\\}

\maketitle

\begin{abstract}
\normalsize We study competition between s-wave order and d-wave order through two holographic superconductor models. We find that once the coexisting phase appears, it is always thermodynamically favored, and that the coexistence phase is narrow and one condensate tends to kill the other. The phase diagram is constructed for each model in terms of temperature and the ratio of charges of two orders. We further compare the behaviors of some thermodynamic quantities, and discuss the different aspects and identical ones between two models.
\end{abstract}

\tableofcontents

\section{Introduction}
\label{sect:introduction}

One of the most studied subjects in the AdS/CFT correspondence~\cite{Maldacena:1997re,Gubser:1998bc,Witten:1998qj} is holographic superconductors, which may shed light upon real word strongly coupled superconductivity. Some holographic superconductor models with different symmetry of condensation have been constructed, including s-wave~\cite{Hartnoll:2008vx,Gubser:2008px,Hartnoll:2008kx}, p-wave~\cite{Gubser:2008wv,Cai:2013pda,Cai:2013aca,Cai:2013kaa,Cai:2014ija} and d-wave~\cite{Chen:2010mk,Benini:2010pr,Chen:2011ny,Kim:2013oba}. Such holographic setups indeed reveal some basic properties of the real superconductors. Nevertheless, most of studies in the literature focus on the case with only a single order parameter.

On the other hand, there are various orders in real high temperature superconductors~\cite{MgB2-nature,MgB2,Berg:2009dga,Zaanen:2010yk,Fujimoto,Goswami:2013wta}. Thus it is desirable to generalize the single order parameter case to multi order parameter case. Indeed, the holographic framework provides us a convenient way to uncover the interaction among those orders by simply adopting dual dynamical fields in the bulk with appropriate couplings.
Following this strategy, several attempts on the competition of multi order parameters in the holographic superconductor models have already been made. In refs.~\cite{Basu:2010fa,Cai:2013wma}, the authors considered the case of two competing scalar order parameters coupled to one U(1) gauge field in the bulk. They found the signature of a coexisting phase where both scalar order parameters appear at the same time. Another holographic superconductor model with a scalar triplet charged under a SU(2) gauge field in the bulk was built in ref.~\cite{Nie:2013sda}. They showed that the s+p coexisting phase turns out to be thermodynamically favored once it appears. Other related studies can be found in refs.~\cite{Huang:2011ac,Krikun:2012yj,Musso:2013ija,Nitti:2013xaa,Liu:2013yaa,Amado:2013lia,Donos:2012yu,Wen:2013ufa,Amoretti:2013oia,Donos:2013woa}.  In this paper, we will study the competition mechanism between s-wave order and d-wave order.

There are two acceptable holographic models describing the d-wave condensation in the literature, in which the d-wave order parameter is dual to a charged massive spin two field propagating in an asymptotically AdS background. The authors of ref.~\cite{Chen:2010mk} first constructed a minimal gravitational model by introducing a symmetric, traceless rank-two tensor field minimally coupled to a U(1) gauge field in the background of the AdS black hole. The d-wave condensate appears below a critical temperature via a second order phase transition, resulting in a superconducting phase with no hard gap for its optical conductivity. Let us call it CKMWY d-wave model in terms of the initials of the five authors. The other holographic d-wave model was proposed soon after the first one with the same matter fields but much more complex interactions~\cite{Benini:2010pr}. The phase diagram, optical conductivity, as well as fermion spectral function were investigated in detail.
With a fixed gravity background, this model has advantages such as being ghost-free and having the right propagating degrees of freedom. This model will be named as BHRY d-wave model in what follows.  To realize s-wave order, we will take advantage of the well known Abelian-Higgs model~\cite{Hartnoll:2008vx} in terms of a complex scalar field charged under a U(1) gauge field in the bulk.

In order to realize the condensation of s-wave order and d-wave order in one holographic model, we can simply combine the Abelian-Higgs model with a  d-wave model.
Thus, we could have  two holographic models with s-wave order and d-wave order.  Actually, we will study the competition between s-wave order and d-wave order for both cases in the probe limit where one neglects the back reaction of matter fields to the background geometry. The phase structures are given and the behaviors of the thermodynamic quantities for the s+d coexisting phase are also studied. The coexisting phase does appear in both models and is thermodynamically favored. Apart from the above common features, the behavior of the ratio of superconducting charge density over the total charge density versus temperature in two models is different.
We also analyze the optical conductivity of the coexisting phase and find some new features.~\footnote{While this work was being prepared, the paper \cite{Nishida:2014lta} appeared in arXiv, which discussed the s+d order coexisting phase, based on the d-wave model proposed in ref.~\cite{Benini:2010pr}, by introducing a coupling between the scalar field and the tensor field, and studied the phase structure in terms of the coupling parameter and temperature with fixed charges of two orders. In our discussion, there is no direct interaction between scalar and tensor fields and our model parameter is the ratio of two fields.  Note that in paper \cite{Nishida:2014lta}, when the coupling $\eta=0$, there also exists coexisting phase under the model parameters $m_1^2=-2$, $m_2^2=0$ and $q_2=1.95$. Both results are consistent with each other in that case.}

The paper is organized as follows. First we study the competition mechanism in the s-wave + BHRY d-wave model in Section~\ref{sect:superfluid}, by investigating including the phase transition, thermodynamics and optical conductivity. We discuss the competition between two orders for the s-wave + CKMWY d-wave model in Section~\ref{sect:superconductor}. We will also give a comparison between the two models. Conclusions and discussions are given in Section~\ref{sect:conclusions}.


\section{The s-wave + BHRY d-wave model}
\label{sect:superfluid}
To study the competition between s-wave and d-wave orders, let us first start with the holographic model by combining the  Abelian-Higgs s-wave model~\cite{Hartnoll:2008vx} and BHRY d-wave model~\cite{Benini:2010pr}. The holographic model with a scalar field $\psi_1$, a symmetric tensor field $\varphi_{\mu\nu}$ and a U(1) gauge field $A_\mu$ is described by the following action:
\begin{equation}\label{BHRY}
\begin{split}
S=\frac{1}{2\kappa^2}\int d^{4}x\sqrt{-g}(- \frac{1}{4} F_{\mu\nu} F^{\mu\nu} -|D\psi_1|^2-m_1^2|\psi_1|^2+ \mathcal{L}_d),\\
\mathcal{L}_d=-|\tilde{D}_\rho\varphi_{\mu\nu}|^2+2|\tilde{D}_\mu\varphi^{\mu\nu}|^2+
|\tilde{D}_\mu\varphi|^2
-\big[\tilde{D}_\mu\varphi^{*\mu\nu}\tilde{D}_\nu \varphi+\text{h.c.}\big]-i q_2 F_{\mu\nu} \varphi^{*\mu\lambda} \varphi^\nu_\lambda\\
-m_2^2\big(|\varphi_{\mu\nu}|^2-|\varphi|^2\big)+2R_{\mu\nu\rho\lambda} \varphi^{*\mu\rho}\varphi^{\nu\lambda}
-R_{\mu\nu}\varphi^{*\mu\lambda}\varphi^\nu_\lambda-\frac{1}{4} R |\varphi|^2,
\end{split}
\end{equation}
where $D_{\mu} = \nabla_\mu - i q_1 A_\mu$ and $\tilde{D}_\mu = \nabla_\mu - i q_2 A_\mu$, $\varphi\equiv\varphi_\mu^\mu$, $\varphi_\rho\equiv G^{\mu\lambda}\tilde{D}_{\lambda}\varphi_{\mu\rho}$ and ${R^\mu}_{\nu\rho\lambda}$ is the Riemann tensor of the background metric. $\psi_1$ is the scalar order and $\psi_{\mu\nu}$ is the tensor order. The parameters $q_1$ and $q_2$ are the charges of the scalar and the tensor fields, respectively. One can perform a rescaling to set the charge $q_1$ of the scalar to be unity. Then the phase structure of this theory is determined by the ratio $q_2/q_1$ by fixing the mass square of the scalar field $m_1^2$ and the mass square of the tensor field $m_2^2$. We shall set $q_1=1$ without loss of generality in the following discussion.

The corresponding equations of motion are as follows,
\begin{eqnarray}
\label{EOM probe action}
&& 0 =g^{\mu\nu} D_{\mu} D_{\nu} \psi_1 - m_1^2 \psi_1, \\
&& 0 = (\nabla_\alpha\nabla^\alpha - m_2^2) \varphi_{\mu\nu} - 2 \tilde{D}_{(\mu} \varphi_{\nu)} + \tilde{D}_{(\mu} \tilde{D}_{\nu)} \varphi - g_{\mu\nu} \big[ (\nabla_\alpha\nabla^\alpha - m_2^2) \varphi - g^{\rho\lambda}\tilde{D}_{\lambda} \varphi_\rho \big] \\
&&\qquad + 2 R_{\mu\rho\nu\lambda} \varphi^{\rho\lambda} - g_{\mu\nu} \frac {R}{4} \varphi - i\frac{q_2}{2} \big( F_{\mu\rho} \varphi^\rho_\nu + F_{\nu\rho} \varphi^\rho_\mu \big), \\
&& \nabla_\mu F^{\mu\nu} = J^\nu,
\end{eqnarray}
where
\begin{equation}
J^\nu =i q_1 \psi_1^{*}g^{\mu\nu} D_{\mu}\psi_1 + i q_2 \varphi^*_{\alpha\beta} (g^{\mu\nu} \tilde{D}_{\mu} \varphi^{\alpha\beta} - g^{\alpha\lambda} \tilde{D}_{\lambda} \varphi^{\nu\beta}) + i q_2 (\varphi^*_\alpha - \tilde{D}_\alpha \varphi^*)(\varphi^{\nu\alpha} - g^{\nu\alpha} \varphi) + \text{h.c.} \;.
\end{equation}
 Note that here there is no direct interaction between $\psi_1$ and $\varphi_{\mu\nu}$, but they interact with each other via the U(1) gauge field and the strength is controlled by the ratio of charge $q_2/q_1=q_2$.

\subsection{The ansatz and equations of motion}
Working in the probe limit, we choose the background metric to be the 3+1 dimensional AdS-Schwarzschild black hole with planar horizon, which reads
\begin{equation}
\label{metric}
ds^2=-f(r)dt^2+\frac{dr^2}{f(r)}+r^2(dx^2+dy^2),
\end{equation}
where $f(r)=r^2-\frac{r_h^3}{r}$ and the AdS radius has been set to be unity. The horizon is located at $r_h$ and the Hawking temperature for this black hole is $T=\frac{3r_h}{4\pi}$, which is also the temperature of the dual field theory.

We consider an ansatz where $\varphi_{\mu\nu}$ and $A_\mu$ depend only on the radial coordinate $r$ and the spatial components of $\varphi_{\mu\nu}$ are turned on only. According to ref. \cite{Benini:2010pr}, it is consistent to turn on a single component of $\varphi_{\mu\nu}$ and to set other components of the gauge field except for $A_t$ to be zero. Then our ansatz is
\begin{equation}
\label{ansatz_1}
A_\mu \, dx^\mu =  \phi(r) \, dt  \;, \quad \psi_1=\psi_1(r)\quad
\varphi_{xy}=\varphi_{yx} = \frac{r^2}{2} \, \psi_2(r) \;,
\end{equation}
with $\phi(r)$, $\psi_1(r)$ and $\psi_2(r)$ all real functions.

With the above ansatz~\eqref{ansatz_1}, the equations of motion for $\phi$, $\psi_1$ and $\psi_2$ are given by
\begin{eqnarray}\label{EOMs}
\begin{split}
\phi'' + \frac{2 \phi'}{r} - \frac{2}{f}\phi \psi_1^2- \frac{q_2^2} {f} \phi \psi_2^2=&0, \\
\psi_1'' + \frac{f'}{f} \psi_1'+ \frac{2}{r} \psi_1' + \frac{\phi^2}{f^2} \psi_1- \frac{m_1^2}{f} \psi_1=&0, \\
\psi_2'' + \frac{f'}{f} \psi_2' + \frac{2}{r} \psi_2' + \frac{ q_2^2 \phi^2}{f^2} \psi_2- \frac{m_2^2}{f} \psi_2=&0.
\end{split}
\end{eqnarray}
Here the prime denotes the derivative with respect to $r$. From the above explicit equations of motion, we can easily get the s-wave or d-wave superconductivity by turning off the tensor degree of freedom $\psi_2$ or the scalar field $\psi_1$, respectively. Therefore, with this model at hand, we can study the competition mechanism between the s-wave order and d-wave order.

It is easy to see that equations~\eqref{EOMs} has a symmetry
\begin{equation}\label{symmetry}
m_1^2\leftrightarrow m_2^2,\ q_2\rightarrow 1/q_2, \ \phi\rightarrow q_2 \phi,\  \psi_1\rightarrow q_2\psi_2 /\sqrt{2},\ \ \psi_2\rightarrow\sqrt{2} q_2\psi_1.
\end{equation}
Under this symmetry transformation, the role of s-wave and d-wave would interchange each other. Without loss of generality, here we  focus on the case $m_1^2<m_2^2$.

In order to find the solutions for all the three functions $\mathcal{F}=\{\phi, \psi_1,\psi_2\}$, one must specify suitable boundary conditions both at the AdS boundary and
at the horizon. We demand that the matter fields near the boundary $r\rightarrow \infty$ should behave as
\begin{equation}
\label{boundary condition_1}
\phi=\mu-\frac{\rho}{r}+\cdot\cdot\cdot, \quad \psi_1=\frac{\psi_{1+}}{r^{\Delta_{1+}}}+\cdot\cdot\cdot,\quad \psi_2=\frac{\psi_{2+}}{r^{\Delta_{2+}}}+\cdot\cdot\cdot,
\end{equation}
where $\Delta_{1+}=\frac{3+\sqrt{9+4m_1^2}}{2}$ and $\Delta_{2+}=\frac{3+\sqrt{9+4m_2^2}}{2}$.~\footnote{Following ref.~\cite{Benini:2010pr}, the unitary bound implies that $\Delta_{2+}\geq 3$ for spin two operators. Therefore, the mass of $\varphi_{\mu\nu}$ has a lower bound, i.e., $m_2^2\geq 0$.}
Note that the fall-off of $\psi_1$ and $\psi_2$ is chosen so that the dual charged operators have no deformation but can acquire expectation value spontaneously.
According to the holographic dictionary, up to a normalization, the coefficients $\mu$, $\rho$, $\psi_{1+}$ and $\psi_{2+}$ are interpreted as chemical potential, charge density, the expectation values of scalar operator $\mathcal{O}_{1}$ and the spin two operator $\mathcal{O}_{xy}$, respectively.

At the horizon, in addition to $f(r_h)=0$, one must require $\phi(r_h)=0$ in order that $g^{\mu\nu}A_{\mu}A_{\nu}$ is finite at the horizon. Regularity of the solution at the horizon $r=r_h$ requires that all the functions have finite value and admit a series expansion in terms of $(r-r_h)$ as
\begin{equation}
\label{expansion}
\mathcal{F}=\mathcal{F}(r_h)+\mathcal{F}'(r_h)(r-r_h)+\cdot\cdot\cdot.
\end{equation}
By plugging the expansion~\eqref{expansion} into~\eqref{EOMs}, one can find that there are four independent parameters at the horizon $\{r_h,\psi_1(r_h),\psi_2(r_h),\phi'(r_h)\}$. Note that the equations of motion~\eqref{EOMs} have a useful scaling symmetry
\begin{equation}
r\rightarrow\lambda r,\ \ (t,x,y)\rightarrow\lambda^{-1}(t,x,y),\ \  f\rightarrow\lambda^2 f,\ \ \ \phi\rightarrow\lambda\phi,
\end{equation}
where $\lambda$ is a real positive constant.
Taking advantage of the above scaling symmetry, we can set $r_h=1$ for performing numerics. Then we have three independent parameters $\{\psi_1(r_h),\psi_2(r_h),\phi'(r_h)\}$, where two of them will be chosen as shooting parameters to match the asymptotic expansion \eqref{boundary condition_1}. After solving the set of equations, we can obtain the condensates $\langle \mathcal{O}_1\rangle$ and $\langle \mathcal{O}_{xy}\rangle$, chemical potential $\mu$ and charge density $\rho$ by reading off the corresponding coefficients in \eqref{boundary condition_1}, respectively.

The normal phase in the dual field theory is characterized by the vanishing vacuum expectation values of both condensates, which corresponds to vanishing scalar field $\psi_1$ and spin two tensor field $\psi_2$ in the bulk. The gravity background describing the normal phase can be solved exactly, which reads
\begin{equation}
\phi=\mu(1-\frac{r_h}{r}),\quad\psi_1(r)=\psi_2(r)=0.
\end{equation}
\subsection{Qualitative analysis}
Before solving the set of coupled equations~\eqref{EOMs} numerically, we make a briefly qualitative analysis on the possible phases for such a model. Following ref.~\cite{Basu:2010fa}, we rephrase the equations for the s-wave and d-wave as a potential problem. It is convenient to work in the z-coordinate where $z=1/r$. In this coordinate, the infinite boundary is now at $z=0$, while the horizon is at $z=1/r_h = 1$.
With the transformation $\tilde{\psi_1}=\psi_1 z^{-1}$ and $\tilde{\psi_2}=\psi_2 z^{-1}$, the evolution equations for s-wave and d-wave in equations~\eqref{EOMs} can be rewritten as follows
\begin{eqnarray}\label{potential_form}
z^{2}f(z^2f\tilde{\psi_1}_{,z})_{,z}-V_{1eff}\tilde{\psi_1}&=&0,\nonumber\\
z^{2}f(z^2f\tilde{\psi_2}_{,z})_{,z}-V_{2eff}\tilde{\psi_2}&=&0,
\end{eqnarray}
where $V_{1eff}(z)=-f^2(\frac{\phi^2}{f^2}-\frac{m_1^2}{f}+\frac{f_{,z}}{f}z^3)$ and $V_{2eff}(z)=-f^2(\frac{q_2^2 \phi^2}{f^2}-\frac{m_2^2}{f}+\frac{f_{,z}}{f}z^3)$.

After introducing a new variable $y$, the above equations~\eqref{potential_form} can be further expressed as
\begin{eqnarray}
\frac{d^2}{d y^2}\tilde{\psi_1}-\tilde{V}_{1eff}(y)\tilde{\psi_1}&=&0,\nonumber \\
\frac{d^2}{d y^2}\tilde{\psi_2}-\tilde{V}_{2eff}(y)\tilde{\psi_2}&=&0,
\end{eqnarray}
where $dy=-\frac{dz}{z^2 f}$ with $y\rightarrow \infty$ as $z\rightarrow 1$ and $y\rightarrow 0$ as $z\rightarrow 0$. Now in terms of the new variable $y$, the equations of motion for s-wave and d-wave are rephrased as a potential problem on a semi infinite line, i.e., $y\in [0, \infty)$. We will analyze this potential problem in detail case by case.

Our discussion is base upon the lemma proven in ref.~\cite{Basu:2010fa}: For two potentials $V_1$ and $V_2$ over the same domain with $V_1>V_2$, the lowest eigenvalue of $V_1$ would be strictly greater than the lowest eigenvalue of $V_2$. The lemma implies that if the lowest eigenvalue mode for $V_2$ is a zero mode and then $V_1$ can not have a bound state or a zero mode.
Note  that we focus on the case $m_1^2<m_2^2$ and have set $q_1=1$.

\subsubsection{$q_2^2<1$ case}
In this case, no matter which gauge field configuration we choose, we always have $V_{1eff}<V_{2eff}$, which implies that a zero mode of s-wave should form before a zero mode of d-wave. Although the condensate of s-wave changes the gauge field profile, according to the lemma, with the modified gauge potential $V_{1eff}<V_{2eff}$ and a s-wave has a node-less condensate, no zero mode or bound state of d-wave exists. That is to say, in the phase with s-wave condensed yet, the d-wave can not condense. Therefore the phase structure of the system is the same as that of s-wave holographic superconductor with a single scalar.
\subsubsection{$q_2^2\geq1$ case}
This case is much more complicated. One may expect that the d-wave field with large charge $q_2$ will always dominate. However, the potential $V_{1eff}$ diverges like $\frac{1}{y^2}$ near the boundary $y=0$ when we lower the temperature~\footnote{Note that in the holographic model, only the ratio $\mu/T$ matters. Lowering the temperature is equivalent to increasing the chemical potential. Here we choose to vary the temperature and keep the chemical potential fixed through the whole paper.}. Therefore, lowering the temperature possibly makes the mass dependent potential more important and hence the s-wave tends to dominate. We will confirm this with the numerical calculation.

\subsection{Thermodynamics and phase transition}
Our main purpose is to observe the phase diagram of the model in terms of temperature and  the charge of the tensor field $q_2$.~\footnote{Note that we have set the charge of the scalar field to be unity. Therefore it is better to view $q_2$ as the ratio of charges between the tensor field and the scalar field. } We set the mass square $m_1^2=-2$ and $m_2^2=7/4$ in this paper.  We expect that the model would admit three different superconducting phases. The first superconducting phase corresponds to the pure s-wave with $\psi_1\neq 0$ and $\psi_2=0$. The second one is the pure d-wave with $\psi_2\neq 0$ and $\psi_1=0$. The third superconducting phase admits the coexisting  of the s-wave and d-wave orders.

\begin{figure}[h!]
\centering
\includegraphics[scale=0.85]{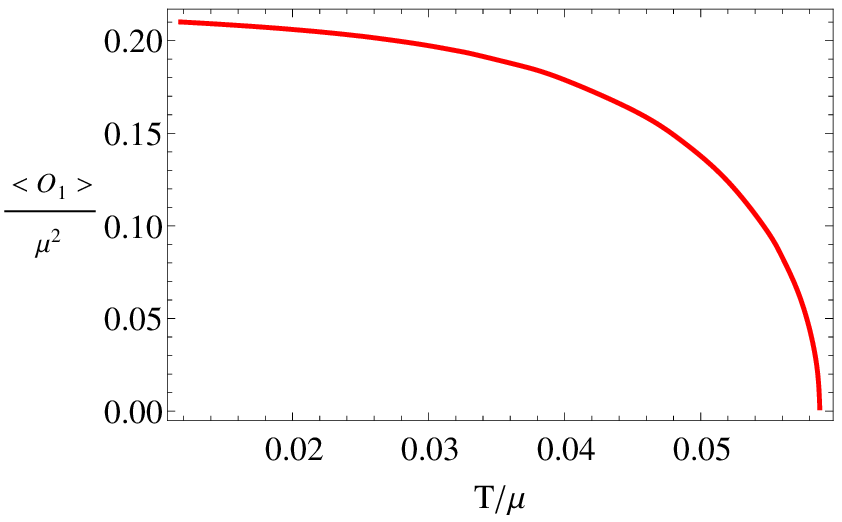}\ \ \ \
\includegraphics[scale=0.85]{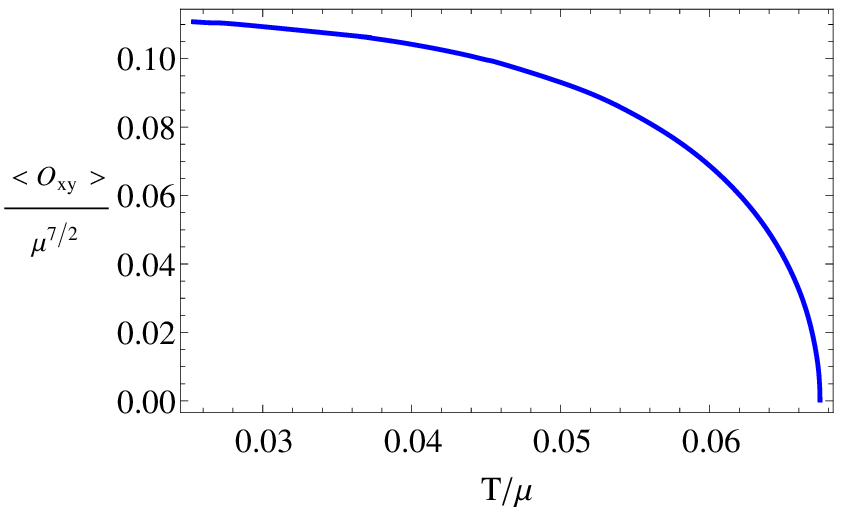}\ \ \ \
 \caption{\label{sd}The left plot shows the condensate as a function of temperature for the pure s-wave phase and the right plot is for the condensate of the pure d-wave phase. When one lowers the temperature, the s-wave order or d-wave order emerges at a critical temperature.}
\end{figure}

Here we take $q_2=2.66$ as a typical example. The condensations for pure s-wave and pure d-wave superconducting phases are depicted in figure~\ref{sd}. As we lower the temperature, the normal phase becomes unstable to developing scalar/tensor hair at a certain critical temperature $T_c$.

For the given charge, one can see that the critical temperature of pure s-wave is lower than the one for the d-wave case. Thus when we lower the temperature, the d-wave order phase should first appear. Once the d-wave order appears, if one goes on lowering the temperature, an interesting question arises: whether the other condensate happens or not?

\begin{figure}[h!]
\centering
\includegraphics[scale=1]{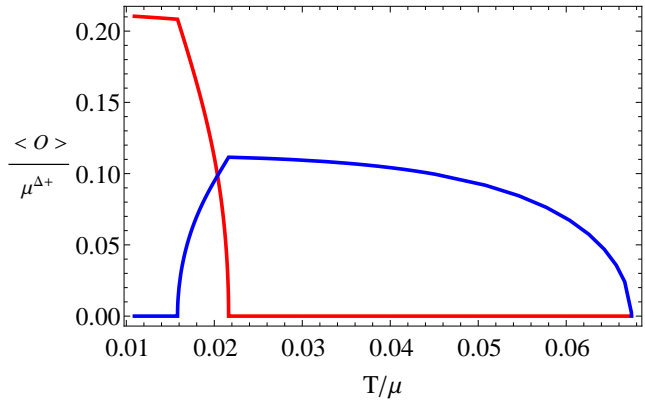}\ \ \ \
 \caption{\label{co} Condensate of the operators in the s+d coexisting phase. The blue curve is for the condensate of the d-wave operator, while the red curve is for the s-wave operator. We see that the d-wave order first condenses, then both orders coexist, finally the d-wave order disappears, leaving only the s-wave order.}
\end{figure}

Our numerical results confirm that the model does admit the coexistence region of two orders with different symmetry, which is drawn in figure~\ref{co}. We can see that as one lowers the temperature, the d-wave order first condenses at $T_c$ where the superconducting phase transition happens. When we continue lowering the temperature to a certain value, say $T_c^{sd1}$, the s-wave order begins to condense, while the condensate of d-wave order decreases, resulting in the state with both orders; if one further lowers the temperature, the d-wave condensate quickly goes to zero at a temperature at $T_c^{sd2}$. When temperature is lower than $T_c^{sd2}$, there exists only the s-wave order. The coexisting phase with both s-wave order and d-wave order can only appear in a narrow range $T_c^{sd2}<T<T_c^{sd1}$.

Based on the above discussion, we have totally three different superconducting phases in our model. In order to determine which phase is thermodynamically favored, we should compare the free energy of the system for each phase. Here we will work in grand canonical ensemble, where the chemical potential is fixed. In the gauge/gravity duality the grand potential $\Omega$ of the boundary thermal state is identified with temperature times the on-shell bulk action with Euclidean signature. Because we work in the probe limit, we only need to consider the contribution from the matter fields to the free energy. The Gibbs free energy can be expressed as
\begin{eqnarray}
\label{Gibbs}
\frac{2\kappa^2 \Omega}{V_2}&=&-\frac{1}{2}\mu\rho-\int_{r_h}^{\infty}dr \sqrt{-g}\frac{1}{2}A_{\nu}(\nabla_{\mu}F^{\mu\nu})\nonumber\\
&=&-\frac{1}{2}\mu\rho-\int_{r_h}^{\infty}dr \frac{1}{2}r^2\phi(-\frac{2\phi'}{r}-\phi''),
\end{eqnarray}
where $V_2=\int dxdy$.

\begin{figure}[h!]
\centering
\includegraphics[scale=0.85]{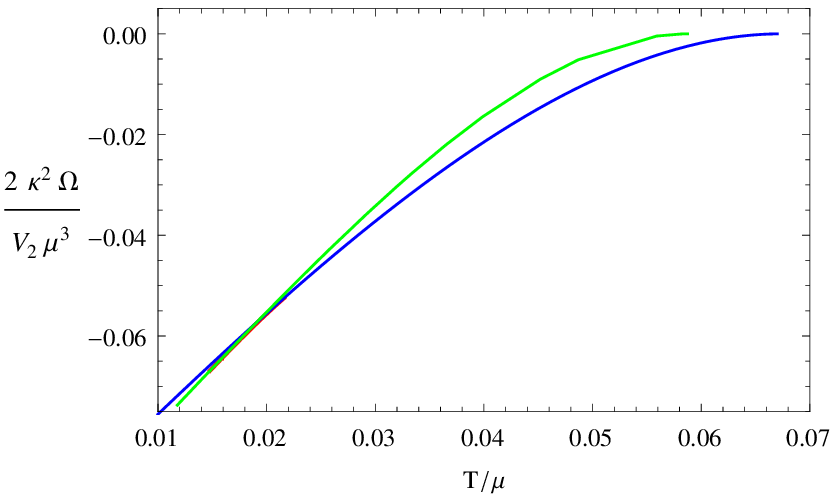}\ \ \ \
\includegraphics[scale=0.85]{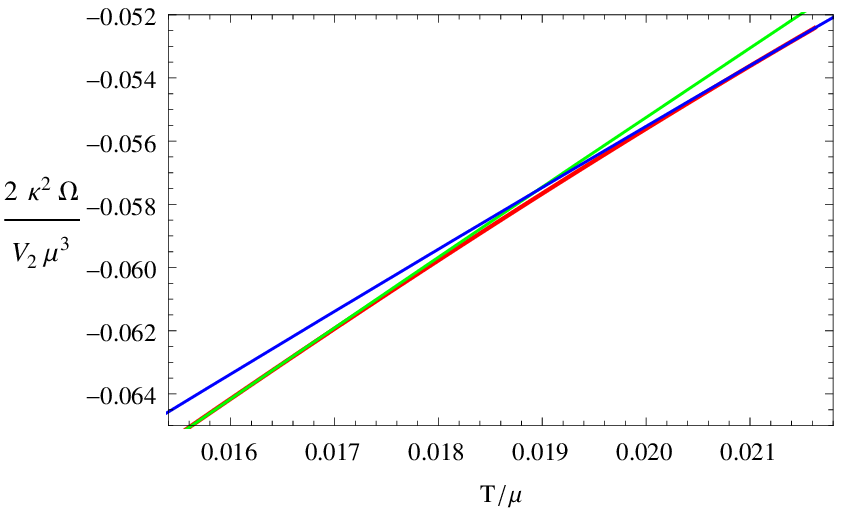}\ \ \ \
 \caption{\label{free} The left plot shows the difference of Gibbs free energy  between the superconducting phase and the normal phase. The blue curve is for the d-wave phase, the green line is for the s-wave phase, while the red curve is for the s+d coexisting phase.  The right plot is an enlarged version of the left one to show the s+d phase more clearly.}
\end{figure}

We plot the difference of the Gibbs free energy between the superconducting phase and the normal phase in figure~\ref{free}. The green curve is shown for the pure s-wave phase, and the blue curve represents the pure d-wave phase. The free energy for the d-wave phase is lower when $T>T_c^{sd1}$, while the free energy for the s-wave is lower when $T<T_c^{sd2}$. When $T_{c}^{sd1}<T<T_c^{sd2}$, the s+d coexisting phase has the lowest free energy, indicating that once the s+d phase exists, it is thermodynamically favored. As we know, there is only a small window admitting the two orders to coexist. Outside the region, it reduces to phases with only a single order. This means that the system is dominated by the d-wave order when $T>T_c^{sd1}$, while dominated by the s-wave order when $T<T_c^{sd2}$. When $T_{c}^{sd1}<T<T_c^{sd2}$, the $s+d$ coexisting phase dominates.

\begin{figure}[h!]
\centering
\includegraphics[scale=0.9]{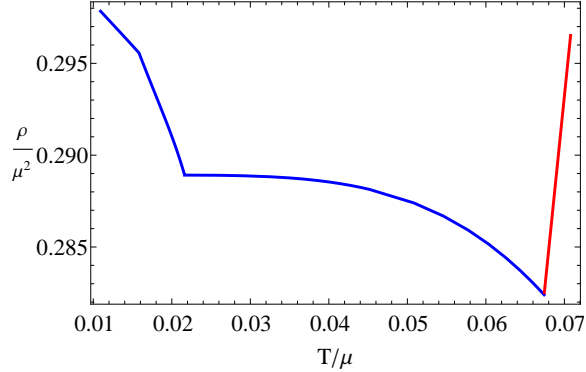}\ \ \ \
 \caption{\label{rho} The total charge density as a function of temperature. The red curve is for the normal phase, while the blue one corresponds to the superconducting phase. There are three special temperatures at which the derivative of charge density with the temperature are discontinuous.}
\end{figure}

\begin{figure}[h!]
\centering
\includegraphics[scale=0.9]{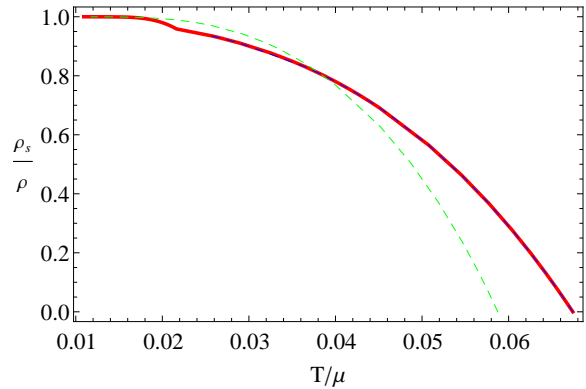}\ \ \ \
 \caption{\label{ratio_1} The ratio of the superconducting charge density over the total charge density $\rho_s/\rho$ versus temperature. The red curve describes the ratio $\rho_s/\rho$ when the system transfers from the d-wave phase to the s-wave phase through the s+d coexisting phase. The green dashed curve is for the ratio $\rho_s/\rho$ of the pure s-wave phase and the blue dashed curve is the ratio for the pure d-wave phase.}
\end{figure}

As we have seen, for suitable $q_2$, the coexisting phase can appear. Once the coexisting solution exists, it is thermodynamically favored, compared to the pure s-wave and pure d-wave phases. Next we give a further investigation on the $s+d$ coexisting phase. We study the behavior of the charge density and the ratio of the superconducting charge over the total charge density $\rho_s/\rho$ with respect to the temperature. Our numerical results are summarized in figure~\ref{rho} and~\ref{ratio_1}.

From figure~\ref{rho}, it can be seen clearly that there exist three particular points at which the derivative of the charge density with respect to temperature is discontinuous, indicating a second order phase transition. The one with the highest temperature is the critical point for the superconducting phase transition, while the remaining two points are inside the superconducting phase, indicating the appearance and disappearance of coexisting phase.

We can also see the signal of phase transition from the behavior of the ratio $\rho_s/\rho$ versus temperature in figure~\ref{ratio_1}. The superconducting charge density $\rho_s$ can be obtained following ref.~\cite{Cai:2012nm}. More precisely, the normal charge density is carried by the black hole and can be read from the electric field at the horizon $\rho_n=\phi'(r_h)$. The total charge density is just $\rho$ in (\ref{boundary condition_1}), determined by the gauge field at the AdS boundary. Thus we can obtain the superconducting charge density $\rho_s=\rho-\rho_n$. The ratio $\rho_s/\rho$ has a small kink in the region for the coexisting phase. When we lower the temperature, the ratio $\rho_s/\rho$ increases in the s+d coexisting phase.

\subsection{Conductivity}
In order to ensure the system is indeed in a superconducting state, and to see whether there are any new phenomena occurring in such coexisting phase, we would like to calculate the optical conductivity $\sigma(\omega)$. To compute the frequency dependent conductivity in the $x$-direction, we consider a set of self-consistent time dependent fluctuations of the fields $A_x$, $\varphi_{ty}$, $\varphi_{ty}^{*}$, $\varphi_{zy}$ and $\varphi_{zy}^{*}$. The coupled linearized algebra-differential equations for the $e^{-i \omega t}$ component of these perturbations are
\begin{subequations}
\label{E:EOMlinear}
\begin{align}
\label{E:EOMAx}
0 &= A_{x,zz} + \frac{f_{,z}}f \, A_{x,z} + \frac{\omega^2}{f^2} \, A_x + \frac{q_2 \psi_2}{2f^2} \, \big[ (\omega - 2q_2 \phi) \varphi_{ty}^* - (\omega + 2q_2\phi) \varphi_{ty} \big] \nonumber \\
&\quad - \frac{iq_2\psi_2}2 \, \big( {\varphi_{zy,z}^*} - \varphi_{zy,z} \big) + \frac{iq_2}{2f} \, (\psi_{2,z} f - \psi_2 f_{,z}) \big( \varphi_{zy}^* - \varphi_{zy} \big) -\frac{2 q_1^2 A_x \psi_1^2}{z^2 f}\ , \\
\label{E:EOMty}
0 &= \varphi_{ty,zz} + \frac2z \, \varphi_{ty,z} - \frac{2f + m_2^2}{z^2 f} \, \varphi_{ty} + \frac{q_2\omega + 2q_2^2 \phi}{4z^2 f} \, \psi_2 A_x 
+ \frac i2\, \big[ 2(\omega + q_2 \phi) \varphi_{zy,z} + q_2 \phi_{,z} \varphi_{zy} \big] \ , \\
\label{E:EOMRzy}
0 &= \big[ (\omega + q_2\phi)^2 z^2 - m_2^2 f \big] \, \varphi_{zy} + \frac i4\,  q_2 f \psi_2 A_{x,z} + \frac i2  q_2 f \psi_{2,z} A_x \nonumber \\
&\quad - i(\omega + q_2\phi) z^2 \varphi_{ty,z} - \frac i2 \, \big[ 4(\omega + q_2 \phi)z + q_2 \phi_{,z} z^2 \big] \, \varphi_{ty} \ ,
\end{align}
\end{subequations}
where we have made a coordinate transformation $z=1/r$ and all quantities in above equations of motion are expressed in terms of coordinate $z$.

The equations for $\varphi_{ty}^*$ and $\varphi_{zy}^*$ are obtained by complex conjugation and an additional transformation $\omega$ to $-\omega$. The functions $\varphi_{zy}$ and $\varphi_{zy}^*$ can be eliminated from the first
two equations using (\ref{E:EOMRzy}), leaving three coupled differential equations for $A_x$, $\varphi_{ty}$ and $\varphi_{ty}^*$.

The boundary conditions we impose on \eqref{E:EOMlinear} are as follows. Since the conductivity is related to the retarded Green's function for the charge current, we should impose the ingoing boundary condition for each fluctuation near the black hole horizon $z_h=1/r_h=1$, i.e., $A_x$, $\varphi_{ty}$ and $\varphi_{ty}^*$ have the behavior as
\begin{equation}
(z_h-z)^{- i \omega / 3} \;.
\label{nearhorizon}
\end{equation}

Near the boundary $z=0$, the asymptotical behavior for the perturbation variables $A_x$, $\varphi_{ty}$ and $\varphi_{ty}^{*}$ is
\begin{eqnarray}
A_x&=&A_x^{(0)}+A_x^{(1)}z+\ldots,\\
\varphi_{ty}&=&\varphi_{ty-}z^{\Delta_-}+\varphi_{ty+}z^{\Delta_+}+\ldots,\\
\varphi_{ty}^{*}&=&\varphi^*_{ty-}z^{\Delta_-}+\varphi^*_{ty+}z^{\Delta_+}+\ldots,
\end{eqnarray}
where $\Delta_{\pm}= \frac{-1\pm \sqrt{9+4m_2^2}}{2}$. Here $\varphi_{ty-}$ and $\varphi^*_{ty-}$ are the sources of the perturbation fields.
After looking for solutions where the source term in the  series expansion of $\varphi_{ty}$ and $\varphi_{ty}^{*}$ vanishes, one can obtain the conductivity as
\begin{equation}
\sigma_{xx}=\frac{A_x^{(1)}}{i \omega A^{(0)}}.
\end{equation}
\begin{figure}[h]
\centering
\includegraphics[scale=0.85]{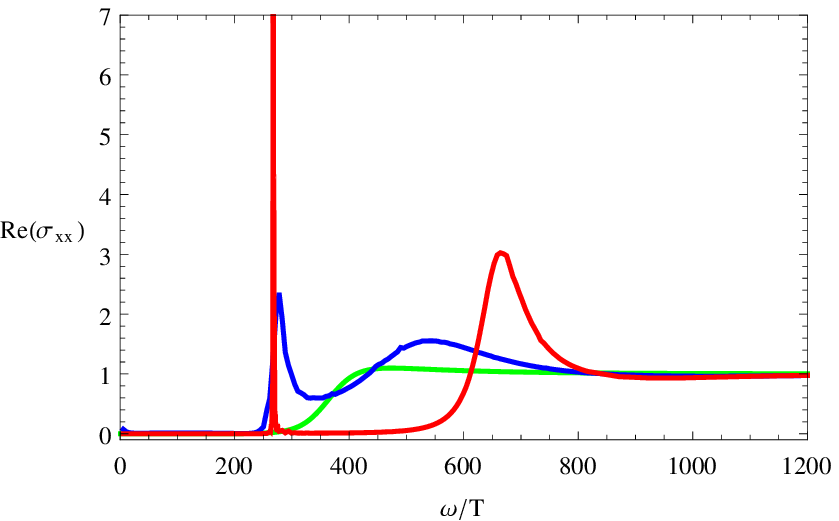}
\includegraphics[scale=0.85]{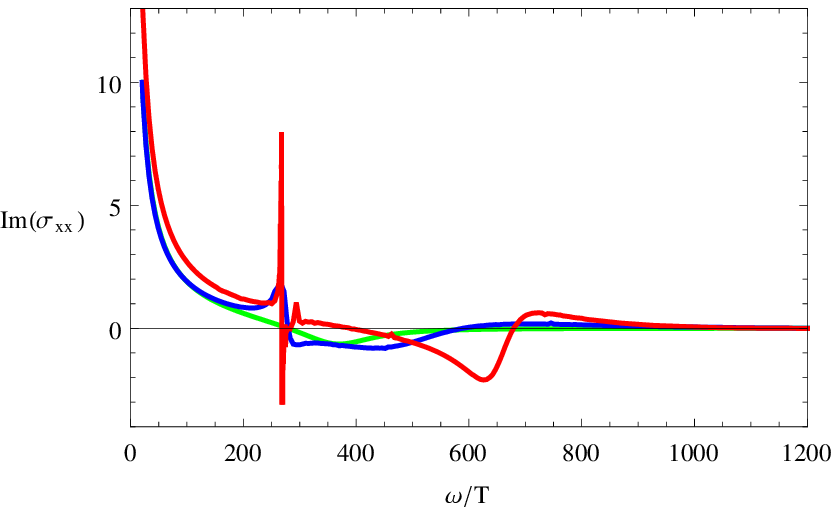}
 \caption{\label{aa} The real part (left) and imaginary part (right) of the conductivity as a function of frequency at temperature $T=0.018\mu$. The red curve is for the s+d coexisting phase, the green line is for the pure s-wave phase and the blue curve for the pure d-wave phase. }
\end{figure}

The numerical results for the  conductivity are shown in figure~\ref{aa}. The green, blue and red curves are for the pure s-wave, pure d-wave and the s+d coexisting phases, respectively. For sufficiently large frequency, Re($\sigma_{xx}$) has a very simple behavior. Much more interesting phenomena happen in the low frequency region. Unlike the s-wave case which only has a bump at $\omega/T \simeq 400$ in figure~\ref{aa}, for pure d-wave condensate, apart from a much more obvious bump at $\omega/T \simeq 500$, Re($\sigma_{xx}$) has an additional spike at a lower frequency.~\footnote{In the case with conformal dimension $\Delta_{2+}=4$ in pure d-wave phase, one can observe two spikes in the conductivity in the small frequency region (see figure~2 in ref.~\cite{Benini:2010pr}). However, for the case with $\Delta_{2+}=7/2$ in this paper, we can only see one spike. Our results suggest that the behavior of conductivity for the BHRY d-wave model~\cite{Benini:2010pr} might depend on the mass of the tensor field. } This spike may indicate the existence of a bound state~\cite{Benini:2010pr}.
One can see clearly that the peak becomes much more sharp in the s+d coexisting state, thus the bound state is enhanced due to the additional condensate of s-wave order.
In addition, the real part of the conductivity has a Direct delta function at $\omega=0$ since the imaginary part of conductivity shown in figure~\ref{aa} has a pole at the origin.

\subsection{Phase diagram}
The calculations of free energy and the conductivity uncover that the coexisting phase is indeed a thermodynamically favored superconducting phase once it appears. However, we do not rule out the possibility that only one order parameter exists for other choices of $q_2$. Thus, it is helpful to construct the phase diagram in terms of temperature $T$ and the charge $q_2$, which tells us in which region the coexisting phase appears.

Let us first make a qualitative discussion from the side of the free energy. Since we fix $m^2_1$ and $m^2_2$ and set $q_1=1$, the critical temperature from the normal phase to the s-wave superconducting phase is fixed, while the critical temperature for the d-wave superconducting transition is proportional to $q_2$. For sufficiently large $q_2$, the critical temperature of the d-wave condensate is higher than the s-wave case, so as one lowers the temperature, the d-wave order will condense first. The s-wave order can only condense behind the d-wave order. However, we have checked that in this case the free energy of the s-wave case is always larger than the one for the d-wave phase. Thus, one can expect that the s-wave order would not dominate the system for very large $q_2$.

On the other hand, for small enough $q_2$, the s-wave order condenses before the d-wave order. As one lowers $q_2$, the critical temperature of d-wave order decreases and the free energy of the d-wave order becomes higher and higher and  will finally be always larger than the one for the s-wave condensate. Therefore, there can be only s-wave order condensation for sufficiently small $q_2$.

For intermediate range of $q_2$, as we show in figure~\ref{free}, the free energy for s-wave case and the one for d-wave case has an intersection at some temperature. If no new phase appears, there should be a first order phase transition from one order to the other order. Nevertheless, the competition between two orders results in the state with both orders coexisting near the crossing point (see figure~\ref{co}).

To summarize, in the $T-q_2$ phase diagram, the phase boundary between the pure s-wave phase and normal phase should be a line parallel to the $q_2$ axis, and the line separating  the pure d-wave phase from normal phase is a straight line passing through the original point $(T,~q_2)=(0,0)$. The s+d coexisting phase can only appear in some region of $q_2$, above which there is only d-wave order, while below which there is only s-wave order. The precise boundary among different phases can only be determined by numerical calculation.

\begin{figure}[h!]
\centering
\includegraphics[scale=1]{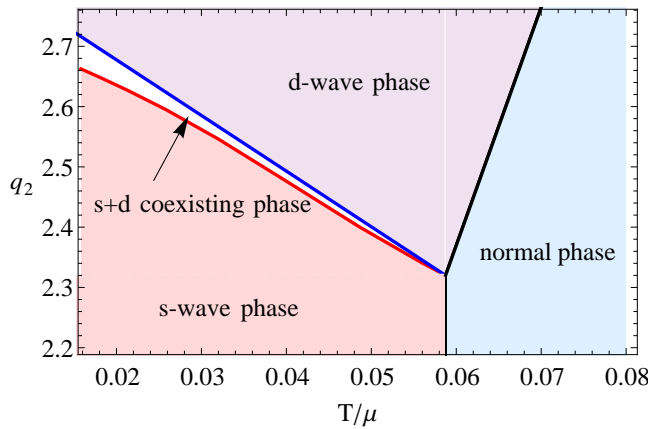}\ \ \ \
 \caption{\label{phase_1} The $q_2$-$T$ phase diagram. The four phases are colored differently and we label the most thermodynamically favored phase in each region.}
\end{figure}

The complete phase diagram with $m_1^2=-2$ and $m_2^2=7/4$ is constructed in figure~\ref{phase_1}. The phase diagram is divided into four parts and the corresponding phase we named in each region is the most thermodynamically favored phase. Indeed, the phase boundary between the normal phase and pure d-wave phase (s-wave phase) is a straight line. The red curve describes the phase transition between the s-wave phase and the s+d coexisting phase. This phase transition occurs when the single s-wave phase becomes unstable to developing a d-wave hair. Therefore, we can derive this red curve using the d-wave as a perturbation on the s-wave superconducting background~\cite{Cai:2013wma}. With the same method, we draw the blue curve which corresponds to the phase transition between $s+d$ coexisting phase and d-wave phase. The full phase diagram is divided into four phases by these curves as boundaries.

From figure~\ref{phase_1}, we see that the coexisting phase exists only in a narrow region in the phase diagram. We denote the critical temperature for a single s-wave or d-wave starting to condense as $T_{cs}$ and $T_{cd}$. If we set the charges of the s-wave and d-wave fields to unity, then $T_{cs}/\mu \simeq 0.0588$ and $T_{cd}/\mu\simeq0.0253$. We see that
\begin{itemize}
\item  In the regime $q_2<T_{cs}/T_{cd}\simeq 2.323$\footnote{It should be noted that when $q_2<T_{cs}/T_{cd}\simeq 2.323$ and $q_2>1.155 T_{cs}/T_{cd}\simeq 2.683$, the $s+d$ coexisting phase does not exist at all. This is different from the case in~ref.~\cite{Nishida:2014lta}. In ref.~\cite{Nishida:2014lta}, for the coupling between the scalar field and the tensor field $\eta=1/10$, the $s+d$ coexisting phase indeed exists, but the free energy of such coexisting phase is larger than those of single order. Thus, from the point of view of thermodynamics, the $s+d$ coexisting phase ceases to exist for $\eta=1/10$ in ref.~\cite{Nishida:2014lta}.}, the s-wave dominates the system and there is no condensation of the d-wave order.
\item As $q_2$ increases beyond $2.323$, the $s+d$ phase appears, which emerges from the d-wave phase. More precisely, as we continue to lower the temperature to $T_c^{sd1}$, the s-wave order begins to condense, while the condensate of the d-wave order decreases, resulting in the phase with both orders; if one further lowers the temperature to $T_{c}^{sd2}$, the d-wave condensate quickly goes to zero; when the temperature is lower than $T_{c}^{sd2}$, there exists only the condensate of s-wave order. There are three second order phase transition as we lower the temperature of the system. The first phase transition happens at $T_c$ when the d-wave order condenses. The second phase transition is at $T_c^{sd1}$ when s-wave order starts condensing. The third one occurs at $T_c^{sd2}$ when the condensation of the d-wave order becomes vanishing.
\item If we continue increasing $q_2$ to the case $q_2>1.155 T_{cs}/T_{cd}\simeq 2.683$, the s-wave order never condenses and the resulting phase diagram is the same as that of model with only d-wave order.
\end{itemize}

Finally, we try to give a qualitative explanation on the mechanism through which the condensation of one order affects the dynamics of the other order~\footnote{We thank the referee for this suggestion.}. Note that here the back reaction is not taken into account. Thus the two fields interact only through their effect on the gauge field once one or both has (have) condensed. Through looking at the gauge field we may give some insight into the competing mechanics between two orders.
\begin{figure}[h!]
\centering
\includegraphics[scale=1]{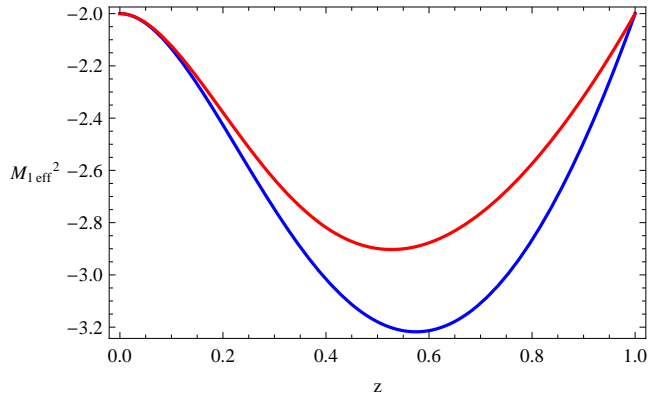}\ \ \ \
 \caption{\label{eff_mass} The blue curve is the effective mass square of s-wave without the condensation of d-wave. The red curve is the effective mass square of the s-wave under the condensation of d-wave. It can be seen clearly that the effective mass of s-wave increases after the condensation of d-wave.}
\end{figure}
\begin{itemize}
\item First, after the d-wave order condenses, if one keeps lowering the temperature and reaches the critical temperature at which the pure s-wave would condense, this condensation does not happen. This is due to the fact that the condensation of the d-wave increases the effective mass of the s-wave, thus prevents the instability of the s-wave to happen, which can be seen from figure~\ref{eff_mass}. This reflects the competition between s-wave and d-wave.
\item However, if we go on decreasing the temperature, the condensation of s-wave does happen. This is due to the fact that the effective mass of the s-wave is lowered and ultimately even if the condensation of the d-wave depleted the gauge potential, the background with only d-wave order becomes unstable.
\item At last, the condensate of the s-wave order kills the first one. This may be thanks to the effective mass of the s-wave being lower.
\end{itemize}
It should be noted that this phenomenon is model dependent. This narrow coexistence region of two superconducting orders and the fact that one condensate can eventually kill the other also happen for two s-wave orders in ref.~\cite{Basu:2010fa} and $p+s$ case in ref.~\cite{Nie:2013sda}. However, it should be noted that this is not the case for the $s+p$ phase studied in ref.~\cite{Amado:2013lia} and the double s-wave scenario in ref.~\cite{Musso:2013ija}, since in both cases the coexisting phase survives even down to a low temperature. Furthermore, the competition diagram here is similar to the competition between the conventional s-wave and the triplet Balian-Werthamer or the B-phase pairings in the doped three dimensional narrow gap semiconductors, such as $\mathrm{Cu}_x\mathrm{Bi}_2\mathrm{Se}_3$ and $\mathrm{Sn}_{1-x}\mathrm{In}_x\mathrm{Te}$ in the condensed matter system~\cite{Pallab:2014}. Although in ref.~\cite{Pallab:2014} the competition is apparently between a s-wave order and a p-wave order, d-wave and p-wave are similar in some circumstances, for example, their excitations of the normal component can be probed using low frequency photons.
\subsection{Generalization to other masses and charges}
\begin{figure}[h!]
\centering
\includegraphics[scale=1]{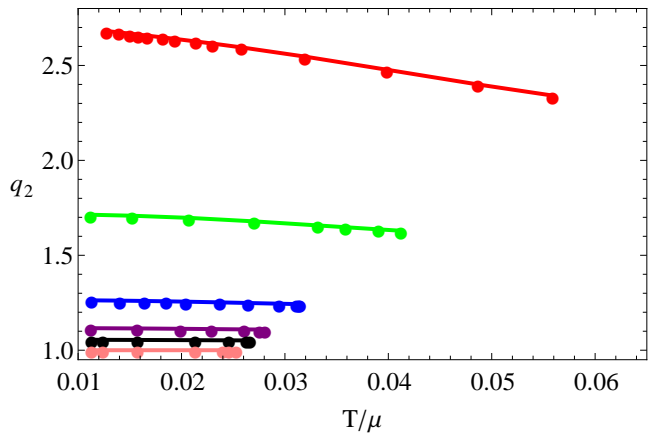}\ \ \ \
 \caption{\label{diff_mass} The parameter space for $s+d$ coexisting phase with d-wave condensed first. Here $m_2^2=7/4$ is fixed and different curves from upper to down correspond to $m_1^2=-2, -5/4, 0, 13/16, 81/64, 7/4$, respectively. For each $m_1^2$, we plot the ratio $q_2$ as a function of the critical temperature below which the d-wave order becomes vanishing.}
\end{figure}

With the same method, we generalize our above analysis to the case with different masses. For convenient, we keep the mass square of the d-wave $m_2^2=7/4$ unchanged. We increase the mass square of $m_1^2$ up to $m_1^2=m_2^2=7/4$. We give the parameter space for the $s+d$ coexisting phase with d-wave condensed first in figure~\ref{diff_mass}.

Figure~\ref{diff_mass} can be derived as follows. Here we want to find the critical ratio $q_2$ such that $T$ is a critical temperature at which the d-wave order $\psi_2$ begins to vanish. At such a temperature, $\psi_2$ is very tiny and can be treated as a perturbation on the background where only $\psi_1$ condenses, i.e.,
\begin{equation}
-\psi_2''-(\frac{f'}{f}+\frac{2}{r})\psi_2'+\frac{m_2^2}{f}\psi_2=\frac{q_2^2 \phi^2}{f^2}\psi_2,
\end{equation}
where the profile of $\phi$ comes from the hairy AdS black hole with only $\psi_1$ condensed. We demand $\psi_2$ to be regular at the horizon and to fall off as in~\eqref{boundary condition_1} near the AdS boundary. Then this equation can be considered as an eigenvalue problem with positive eigenvalue $q_2^2$. The numerical result for the lowest eigenvalue versus temperature is presented in figure~\ref{diff_mass}.

Every point in each curve gives the value of $q_2$ and the corresponding temperature below which the d-wave order tends to vanish. From upper to down, different curves correspond to $m_1^2=-2, -5/4, 0, 13/16, 81/64$, and $7/4$, respectively. We clearly see that as the mass square $m_1^2$ of s-wave increases, the maximal critical temperature brings down. We also find that the value of $q_2$ for a $s+d$ coexisting phase lowers when $m_1^2$ increases. Especially when $m_1^2=m_2^2=7/4$, the value of $q_2$ is always one, which corresponds to the orange line in figure~\ref{diff_mass}. This is due to the symmetry of equations~\eqref{EOMs} mentioned before. With the symmetry, we have $q_2=1/q_1=1$.~\footnote{Figure~\ref{diff_mass} gives the parameter space for the $s+d$ coexisting phase with the d-wave condensed first. By the symmetry~\eqref{symmetry}, we can easily obtain the opposite solution, which is also a coexisting phase but with the s-wave condensed first.}

\section{ The s-wave + CKMWY d-wave model}
\label{sect:superconductor}
With the same strategy, in this section we study the competition between s-wave order and d-wave order in the model combining the Abelian-Higgs s-wave model~\cite{Hartnoll:2008vx} with the CKMWY d-wave model~\cite{Chen:2010mk}. The full action including a U(1) gauge field $A_\mu$, a complex scalar field $\psi_1$ and a symmetric, traceless tensor field $B_{\mu\nu}$ takes the following form
\begin{equation}\label{CKMWY}
S=\frac{1}{2\kappa^2}\int d^4
x\sqrt{-g}(-\frac{1}{4}F_{\mu\nu}F^{\mu\nu}-|D\psi_1|^2-m_1^2|\psi_1|^2+\tilde{\mathcal{L}}_d),
\end{equation}
with
\begin{equation}
\tilde{\mathcal{L}}_d=-g^{\mu\lambda}(\tilde{D}_{\mu}B_{\nu\gamma})^{*}\tilde{D}_{\lambda}B^{\nu\gamma}-m_2^2B_{\mu\nu}^{*}B^{\mu\nu}.
\end{equation}
Here $D_\mu = \nabla_\mu - i q_1 A_\mu$ and $\tilde{D}_\mu = \nabla_\mu - i q_2 A_\mu$.

In the probe limit, matter fields can be treated as perturbations in the 3+1 dimensional AdS black hole background~\eqref{metric}. Let us consider the following ansatz
\begin{equation}
\psi_1=\psi_1(r),\quad B_{xx}=-B_{yy}=\psi_2(r),\quad A_t=\phi(r)dt,
\end{equation}
with all other field components being turned off and $\psi_1(r)$, $\psi_2(r)$ and $\phi(r)$ being real functions. Then the explicit equations of motion are
\begin{equation}\label{EOM_2}
\begin{split}
\phi''+\frac{2}{r}\phi'-\frac{4 q_2^2 \psi_2^2}{r^4f}\phi-\frac{2q_1^2 \psi_1^2}{f}\phi=0,\\
\psi_1''+(\frac{f'}{f}+\frac{2}{r})\psi_1'+\frac{q_1^2\phi^2}{f^2}\psi_1-\frac{m_1^2}{f}\psi_1=0,\\
\psi_2''+(\frac{f'}{f}-\frac{2}{r})\psi_2'+\frac{q_2^2\phi^2}{f^2}\psi_2-\frac{2f'}{rf}\psi_2-\frac{m_2^2}{f}\psi_2=0.
\end{split}
\end{equation}

We use the shooting method to solve the coupled equations of motion~\eqref{EOM_2}. Most of our calculations are the same as those in Section~\ref{sect:superfluid},
for the sake of brevity, we will omit the details about numerical analysis. In this section, we set $q_1=1$, $m_1^2=-2$ and $m_2^2=-\frac{13}{4}$.
The general fall-off of the matter fields near the boundary $r\rightarrow \infty$ behaves as
\begin{eqnarray}
\label{boundary condition}
\phi=\mu-\frac{\rho}{r}+\cdot\cdot\cdot, \ \
\psi_1=\frac{\psi_{1-}}{r^{\Delta_{1-}}}+\frac{\psi_{1+}}{r^{\Delta_{1+}}}+\cdot\cdot\cdot,\ \
\psi_2=\frac{\psi_{2-}}{r^{\Delta_{2-}}}+\frac{\psi_{2+}}{r^{\Delta_{2+}}}+\cdot\cdot\cdot,
\end{eqnarray}
where $\Delta_{1\pm}=\frac{3\pm\sqrt{9+4m_1^2}}{2}$ and $\Delta_{2\pm}=\frac{-1\pm\sqrt{17+4m_2^2}}{2}$. To break the U(1) symmetry spontaneously, we should turn off the source terms, i.e., $\psi_{1-}=\psi_{2-}=0$, then $\psi_{1+}$ and $\psi_{2+}$ are the vacuum expectation values of dual operators, which play the role of order parameters in the boundary field theory.

\subsection{Phase transition and thermodynamics}
So far, the holographic superconducting model with an s-wave order and a d-wave order has been constructed. We are interested in the competition between s-wave and d-wave orders. We will study the phase structure and the behaviors of the thermodynamical quantities. We emphasize the similarity and difference between the model here and the one proposed in the previous section. We take $q_2=1.34$ as an example and summarize our numerical results as follows.

\begin{figure}[h!]
\centering
\includegraphics[scale=0.85]{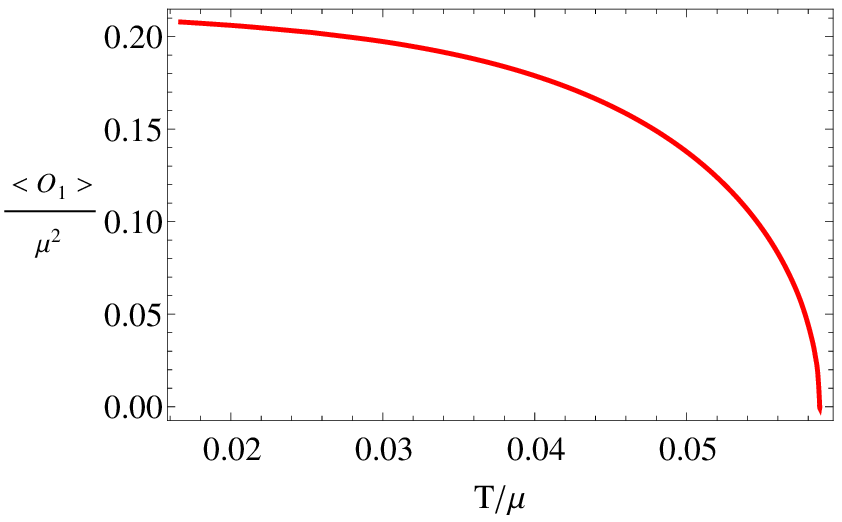}\ \ \ \
\includegraphics[scale=0.85]{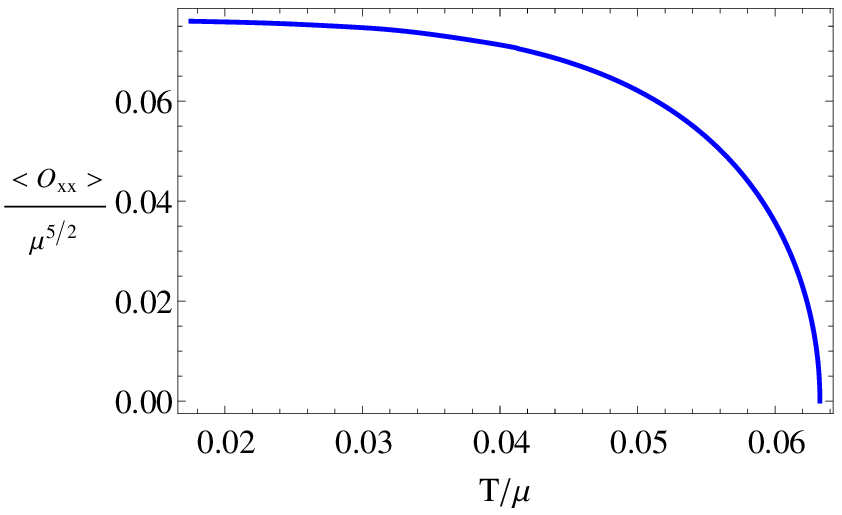}\ \ \ \
 \caption{\label{sandd} The left plot is the s-wave condensate with $\psi_1$, while the right plot is the d-wave condensate with $\psi_2$.}
\end{figure}

First, we investigate all possible phases. As the model in the previous section,  except for the normal phase, there are three additional superconducting phases, the pure s-wave superconducting phase, the pure d-wave superconducting phase and the s+d coexisting phase. We plot the s-wave condensate and the d-wave condensate in figure~\ref{sandd}.

\begin{figure}[h!]
\centering
\includegraphics[scale=0.85]{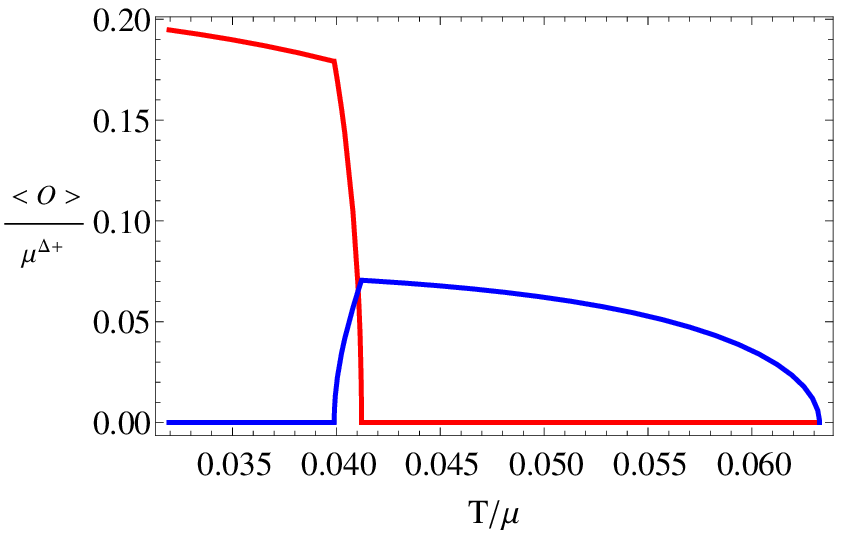}\ \ \ \
\includegraphics[scale=0.85]{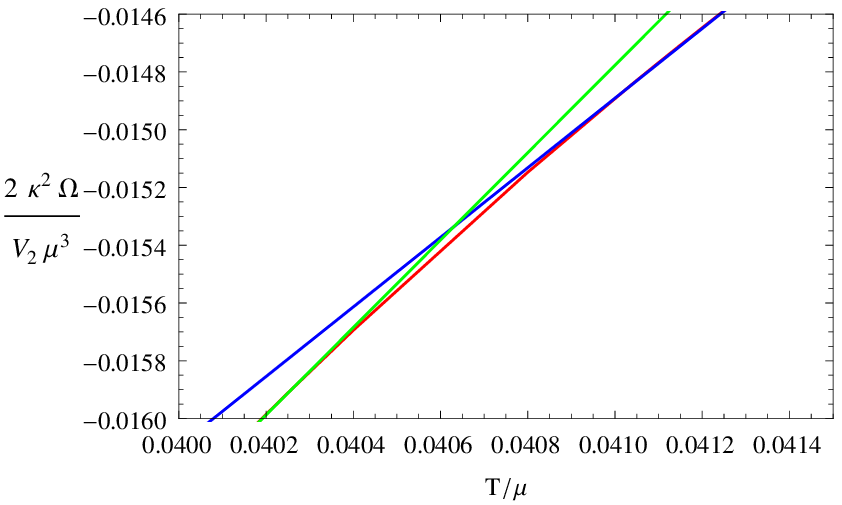}\ \ \ \
 \caption{\label{condensate_2}  The let plot  shows the condensation in the s+d coexisting phase. The right plot shows  the differences of Gibbs free energy between superconducting phases and the normal phase. Here the blue line stands for the d-wave phase, the green one for the s-wave phase and the red one for
the s+d coexisting phase.}
\end{figure}

We also calculate the Gibbs free energy of the model~\eqref{CKMWY} to judge whether the $s+d$ coexisting phase is thermodynamically favored or not. The expression for the free energy turns out to be the same as~\eqref{Gibbs}. We show the condensation of the s+d coexisting phase in the left panel and its Gibbs free energy in the right panel in figure~\ref{condensate_2}. For the case with $q_2\simeq 1.345$, the curves of the free energy for the s-wave phase and d-wave case have an intersection at a temperature, say $T^{cross}$. Since the critical temperature of the d-wave superconducting transition is higher than the one of the s-wave case, the d-wave phase will first appear. We can see that the free energy for the d-wave phase is lower than the s-wave phase when $T>T^{cross}$, while it becomes larger than the s-wave phase when $T<T^{cross}$. One expects that there should be a transition from the d-wave phase to the s-wave phase.

Indeed, as we can see in figure~\ref{condensate_2} a new phase with both s-wave order and d-wave order coexistence can appear near $T^{cross}$. We find that
this s+d coexisting phase has the lowest free energy and is thus thermodynamically preferred to the s-wave phase and d-wave phase. In more detail, as we lower the temperature of the system, it first undergoes a phase transition from the normal phase to the pure d-wave phase at $T_c^d$. Then at $T_c^{sd1}$, a new phase transition occurs, and the system goes into an s+d coexisting phase. At last the system undergoes the third phase transition from the s+d coexisting phase to a pure s-wave phase at $T_c^{sd2}$. Note that all the three phase transitions are second order. The temperature region for the s+d wave coexisting phase is very narrow, which is similar to the previous model in Section~\ref{sect:superfluid}.

\begin{figure}[h]
\centering
\includegraphics[scale=0.9]{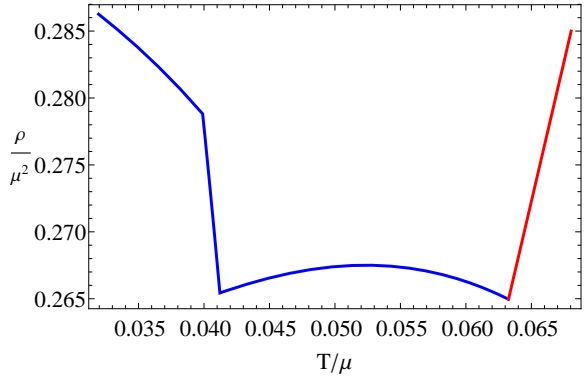}\ \ \ \
\caption{\label{charge_2} The total charge density as a function of the temperature. The red curve is for the normal phase, while the blue one corresponds to the superconducting phase. There are three special temperatures at which the derivatives of charge density with the temperature are discontinuous.}
\end{figure}

The feature of the phase transitions  can also be seen clearly from the charge density as the function of temperature in figure~\ref{charge_2}. We find that the charge density with respect to temperature is continuous, but its derivative is discontinuous at three special points, indicating three second order phase transitions. The first transition from the normal phase to the d-wave superconducting phase occurs at the highest critical temperature. The other two transitions from the d-wave to s+d coexisting phase and from the s+d phase to the s-wave case occur inside the superconducting phase. These features are the same as those for the model in the previous section. But there is a little difference in the behavior of the total charge density for the d-wave phase. In the s-wave + BHRY d-wave model, the total charge density changes monotonously with the temperature, while it behaves non-monotonous in the present model.

\begin{figure}[h]
\centering
\includegraphics[scale=0.9]{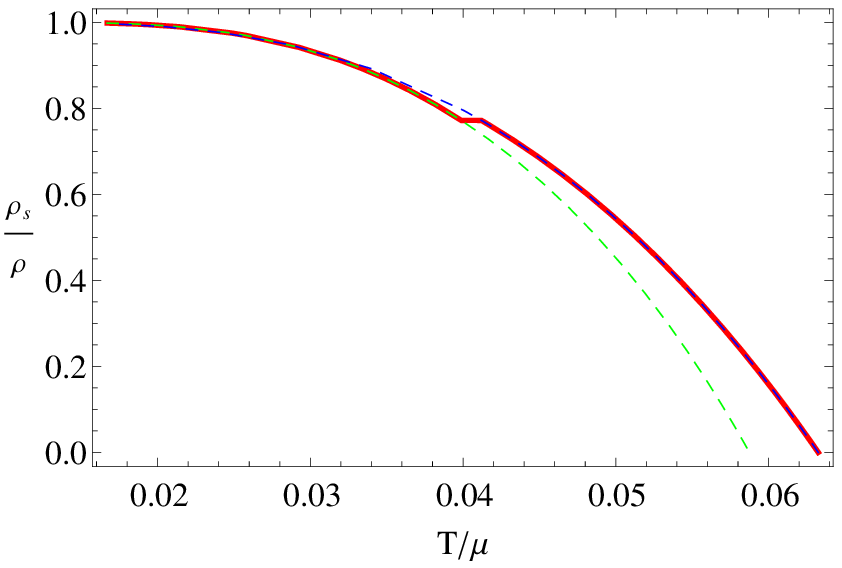}\ \ \ \
\caption{\label{ratio_2} The ratio of the superconducting charge density over the total charge density, $\rho_s/\rho$, with respect to the temperature. The red curve describes the ratio $\rho_s/\rho$ when the system transfers from the d-wave phase to s-wave phase through the s+d coexisting phase. The green dashed blue curve is for the ratio $\rho_s/\rho$ of the pure s-wave phase and the blue dashed curve is the ratio for the pure d-wave phase.}
\end{figure}

The information of the phase transitions can also be revealed via the behavior of the ratio $\rho_s/\rho$ with respect to the temperature.
From figure~\ref{ratio_2}, one can see that the ratio $\rho_s/\rho$ also has a small kink in the region of the coexisting phase. Comparing figure~\ref{ratio_1} with figure~\ref{ratio_2}, we see that in the former case, the green dashed curve for the pure s-wave phase intersects with the blue dashed curve for the pure d-wave phase. In contrast, the green dashed curve in figure~\ref{ratio_2} is always lower than the blue dashed curve. Therefore, as one lowers the temperature, the ratio $\rho_s/\rho$ in the s+d coexisting phase increases for the former \eqref{BHRY}, while it decreases for the latter~\eqref{CKMWY}. The authors of ref.~\cite{Nie:2013sda} investigated an s+p coexisting phase and found the decrease of the ratio $\rho_s/\rho$ in the coexisting phase, similar to figure~\ref{ratio_2}. They suggested that it might be an experimental signal of the phase transition from a single condensate phase to a coexisting phase. Nevertheless, our results uncover that the ratio $\rho_s/\rho$ versus temperature is model dependent.

\subsection{Phase diagram}
\begin{figure}[h!]
\centering
\includegraphics[scale=0.85]{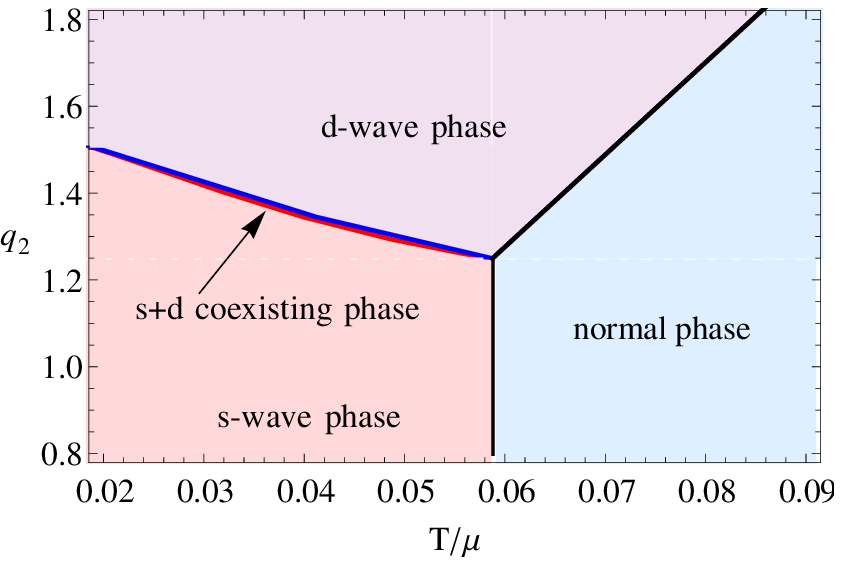}\ \ \ \
\includegraphics[scale=0.85]{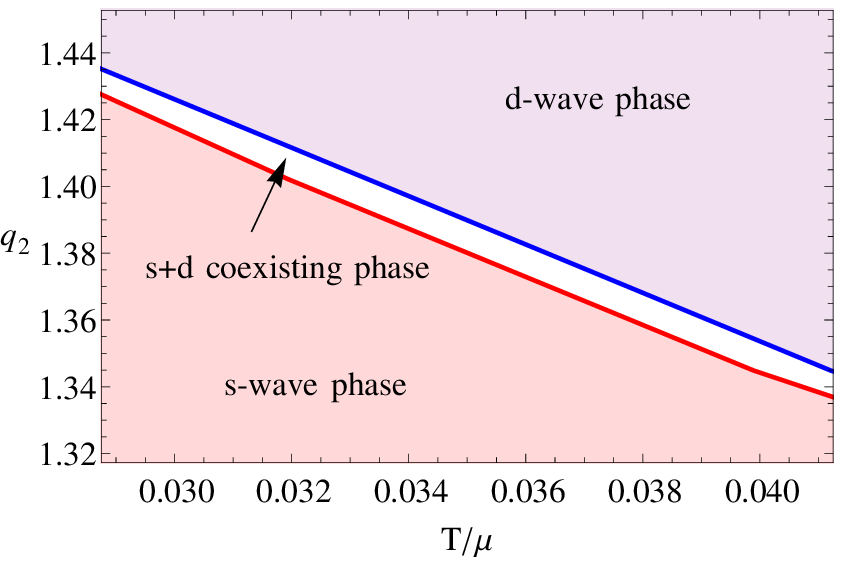}\ \ \ \
 \caption{\label{phase_2} The $q_2$-$T$ phase diagram with $m_1^2=-2$ and $m_2^2=-\frac{13}{4}$. We label the most thermodynamically favored phase in each part. The s+d coexisting phase exists only in a narrow region. The right plot is an enlarged version for the coexisting region in order to see this more clearly.}
\end{figure}

By adopting the same procedure as in section~\ref{sect:superfluid}, we construct the phase diagram for the model~\eqref{CKMWY} with $m_1^2=-2$ and $m_2^2=-\frac{13}{4}$ in the $q_2-T$ plane in figure~\ref{phase_2}. 
As the s-wave + BHRY d-wave model,  the system also contains four kinds of phases known as the normal phase, s-wave phase, d-wave phase and  s+d coexisting phase. The normal phase dominates in the high temperature region, the s-wave phase dominates in the lower temperature region with small $q_2$ below the red curve, and the d-wave phase dominates in the higher temperature zone with large $q_2$ above the blue curve. The s+d coexisting phase is favored in the area between the red and blue curves. The region for the $s+d$ coexisting phase  is very narrow in the phase diagram, which indicates that the s-wave and d-wave phases generally repel each other, but they can coexist in a very small range of temperature.
\subsection{Generalization to other masses and charges}
\begin{figure}[h!]
\centering
\includegraphics[scale=0.85]{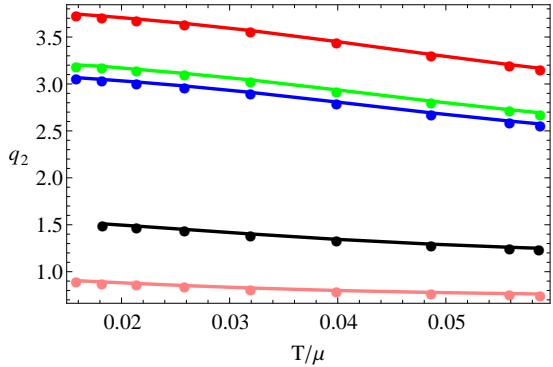}\ \ \ \
 \caption{\label{mass_2} The parameter space for a $s+d$ coexisting phase where d-wave order condenses first. Here we keep $m_1^2=-2$. The curves from upper to down corresponds to $m_2^2=19/4, 2, 89/64, -13/4, -4$, respectively. For each $m_2^2$, we give the ratio $q_2$ as a function of the critical temperature below which the d-wave order tends to vanish.}
\end{figure}

It is clear from the equations~\eqref{EOM_2} that the s-wave and d-wave orders now see different effective potentials. Therefore, the analytical discussion in the previous section in terms of effective potential can give little useful information. We have to resort to numerical methods to find possible solutions. Here we keep $m_1^2$ of the scalar unchanged and increase $m_2^2$ of the d-wave order, which lowers the critical temperature of a single d-wave condensation. We hope to find the coexisting phase with s-wave order condensed before d-wave one by increasing the mass square $m_2^2$. With the same method as done in figure~\ref{diff_mass}, the parameter space for the $s+d$ coexisting phase where d-wave order condenses first is shown in figure~\ref{mass_2}. Nevertheless, from figure~\ref{mass_2} we see that the value of $q_2$ (which can indicate the appearance of $s+d$ coexisting phase) increases when $m_2^2$ is increased. A larger $q_2$ in turn makes the critical temperature of d-wave condensation much more higher. The behavior here is obviously different from the one in previous model (see figure~\ref{diff_mass}). Therefore, it comes as no surprise that we do not find the $s+d$ coexisting solution for which s-wave order condenses before d-wave order.

\section{Conclusions}
\label{sect:conclusions}
The competition between the s-wave condensate and the d-wave condensate has been studied through two holographic models.  The dynamics of s-wave order dual to a complex scalar field is described  by the Abelian-Higgs model. The dynamics of the d-wave order corresponding to a symmetric, traceless spin two tensor field, is determined by the bulk action from BHRY d-wave model~\cite{Benini:2010pr} or  CKMWY d-wave model~\cite{Chen:2010mk}. In our study, we did not include the direct interaction between the scalar field and the tensor field in the bulk, but, they interact with each other through the U(1) gauge field. Note that including a direct interaction between them is equivalent to changing the effective masses of the scalar field and tensor field. Based on these, we give some qualitative explanation on the completing scenario in our cases. Working in the probe limit, we are left with three model parameters, i.e., the mass square $m_1^2$ of scalar field, the mass square $m_2^2$ of tensor field and the charge ratio $q_2/q_1$, where $q_2$ is the charge for the d-wave order and $q_1$ is for the s-wave order. Without loss of generality, we have set $q_1$ to be unity in the numerical calculations.

Based on our analysis, there are similarity and difference between two holographic setups, i.e., the model~\eqref{BHRY} and the model~\eqref{CKMWY}. The common features are as follows:
\begin{itemize}
\item The s+d coexisting phase does exist in a region of the model parameter $q_2/q_1$. Once the coexisting phase appears, it is always thermodynamically favored, compared to the pure s-wave and pure d-wave superconducting phases, which can be seen from the free energy in figure~\ref{free} and figure~\ref{condensate_2}.
\item The phase transition from the coexisting phase to the phase with single order is second order, which can be seen from the charge density versus temperature in figure~\ref{rho} and figure~\ref{charge_2}. In fact, all phase transitions are second order in these two holographic models.
\item One can see from figure~\ref{phase_1} and figure~\ref{phase_2} that the phase structure is very similar for both models. The region for the s+d coexisting phase is very narrow in the phase diagram, indicating that the s-wave and d-wave phases generally repel each other.
\end{itemize}

There exist also  some differences in the two models. Comparing to equations~\eqref{EOMs} and~\eqref{EOM_2}, we see that the first model exhibits an useful symmetry~\eqref{symmetry}. Taking advantage of this symmetry, one can only consider the case with $m_1^2<m_2^2$. Our numerical calculations uncover that, for suitable model parameters, as the temperature is lowered, the s-wave order condenses inside the d-wave order resulting in the coexisting phase with both orders. However, when the scalar order condenses the first one starts to disappear, and finally only the s-wave condensate is left for sufficiently low temperatures. If we change the model parameter $m_1^2\leftrightarrow m_2^2$, the inverse is also true: the condensate of d-wave order emerges following the condensate of s-wave order, and then the d-wave condensate finally kills the s-wave order. Those two kinds of coexisting phase are one to one correspondence.~\footnote{Strictly speaking, this statement is valid for cases with $m_1^2$ and $m_2^2$ both non-negative, since the unitary bound requires the mass square of $\varphi_{\mu\nu}$ should be non-negative.} In contrast, in the second model, we only find the first kind of the coexisting phase. What's more, for the first model~\eqref{BHRY}, the ratio $\rho_s/\rho$ increases in the s+d coexisting phase as the temperature is lowered, while it decreases in the second case~\eqref{CKMWY}. This gives an obvious evidence that the ratio $\rho_s/\rho$ versus temperature is model dependent.

The optical conductivity in the s+d coexisting phase was calculated for the s-wave+BHRY d-wave model~\eqref{BHRY}. We found a remarkable spike in the low frequency region, compared to the case for the pure d-wave superconducting phase, this is due to the additional condensation of the s-wave order in the coexisting phase.

In both models, the s+d coexisting phase is narrow and one condensation tends to kill the other. This is similar to the situation of the coexisting phase with two s-wave orders~\cite{Basu:2010fa} as well as the case with one s-wave order and one p-wave order~\cite{Nie:2013sda}. This competing behavior is similar to the case shown in the condensed matter system~\cite{Pallab:2014}. However, it should note that the competing scenario is model dependent. The cases in ref.~\cite{Musso:2013ija} and ref.~\cite{Amado:2013lia} are different from here. In these two cases, the condensates feed on different charge densities and the coexisting phase survives down to a low temperature.

Note that as found in ref.~\cite{Cai:2013wma}, including the back reaction of matter fields would lead to a much rich phase structure for two s-wave orders model. Therefore, it will be desirable to study a consistent s+d holographic superconducting model with back reaction, although it would be a challenge in some sense due to the complexity of the spin two field theory in curved spacetime.  Nevertheless, we may overcome some difficulty by an effective model with a well-chosen ansatz. We will leave this for further study.

\section*{Acknowledgements}

L.F.L thanks C. P. Herzog, A. Yarom, D. Marity, K. Y. Kim, Z.Y. Fan and Z.Y. Nie for various helps. This work was supported in part by the National Natural Science Foundation of China (No.10821504, No.11035008, No.11375247, No.11205226 and No.41231066), and in part by the Ministry of Science and Technology of China under Grant No.2010CB833004. L.F.L would like to appreciate the Specialized Research Fund for State Key Laboratories of CAS.


\begin{thebibliography}{99}

\bibitem{Maldacena:1997re}
  J.~M.~Maldacena,
  ``The large N limit of superconformal field theories and supergravity,''
  Adv.\ Theor.\ Math.\ Phys.\  {\bf 2}, 231 (1998)
  [Int.\ J.\ Theor.\ Phys.\  {\bf 38}, 1113 (1999)]
  [arXiv:hep-th/9711200].

\bibitem{Gubser:1998bc}
  S.~S.~Gubser, I.~R.~Klebanov and A.~M.~Polyakov,
  ``Gauge theory correlators from non-critical string theory,''
  Phys.\ Lett.\  B {\bf 428}, 105 (1998)
  [arXiv:hep-th/9802109].

\bibitem{Witten:1998qj}
  E.~Witten,
  ``Anti-de Sitter space and holography,''
  Adv.\ Theor.\ Math.\ Phys.\  {\bf 2}, 253 (1998)
  [arXiv:hep-th/9802150].

\bibitem{Hartnoll:2008vx}
  S.~A.~Hartnoll, C.~P.~Herzog and G.~T.~Horowitz,
  ``Building a Holographic Superconductor,''
  Phys.\ Rev.\ Lett.\  {\bf 101}, 031601 (2008)
  [arXiv:0803.3295 [hep-th]].

\bibitem{Gubser:2008px}
  S.~S.~Gubser,
  ``Breaking an Abelian gauge symmetry near a black hole horizon,''  Phys.\ Rev.\ D {\bf 78}, 065034 (2008)  [arXiv:0801.2977 [hep-th]].  



\bibitem{Hartnoll:2008kx}
  S.~A.~Hartnoll, C.~P.~Herzog and G.~T.~Horowitz,
  ``Holographic Superconductors,''
  JHEP {\bf 0812}, 015 (2008)
  [arXiv:0810.1563 [hep-th]].

\bibitem{Gubser:2008wv}
  S.~S.~Gubser and S.~S.~Pufu,
  ``The Gravity dual of a p-wave superconductor,''
  JHEP {\bf 0811}, 033 (2008)
  [arXiv:0805.2960 [hep-th]].

\bibitem{Cai:2013pda}
  R.~-G.~Cai, S.~He, L.~Li and L.~-F.~Li,
  ``A Holographic Study on Vector Condensate Induced by a Magnetic Field,''
  JHEP {\bf 1312}, 036 (2013)
  [arXiv:1309.2098 [hep-th]].

\bibitem{Cai:2013aca}
  R.~-G.~Cai, L.~Li and L.~-F.~Li,
  ``A Holographic P-wave Superconductor Model,''
  JHEP {\bf 1401}, 032 (2014)
  [arXiv:1309.4877 [hep-th]].

\bibitem{Cai:2013kaa}
  R.~-G.~Cai, L.~Li, L.~-F.~Li and Y.~Wu,
  ``Vector Condensate and AdS Soliton Instability Induced by a Magnetic Field,''
  JHEP {\bf 1401}, 045 (2014)
  [arXiv:1311.7578 [hep-th]].

\bibitem{Cai:2014ija}
  R.~-G.~Cai, L.~Li, L.~-F.~Li and R.~-Q.~Yang,
  ``Towards Complete Phase Diagrams of a Holographic P-wave Superconductor Model,''
  JHEP {\bf 1404}, 016 (2014)  [arXiv:1401.3974 [gr-qc]].

\bibitem{Chen:2010mk}
  J.~-W.~Chen, Y.~-J.~Kao, D.~Maity, W.~-Y.~Wen and C.~-P.~Yeh,
  ``Towards A Holographic Model of D-Wave Superconductors,''
  Phys.\ Rev.\ D {\bf 81}, 106008 (2010)
  [arXiv:1003.2991 [hep-th]].

\bibitem{Benini:2010pr}
  F.~Benini, C.~P.~Herzog, R.~Rahman and A.~Yarom,
  ``Gauge gravity duality for d-wave superconductors: prospects and challenges,''
  JHEP {\bf 1011}, 137 (2010)
  [arXiv:1007.1981 [hep-th]].


\bibitem{Chen:2011ny}
  J.~-W.~Chen, Y.~-S.~Liu and D.~Maity,
  ``$d+id$ Holographic Superconductors,''
  JHEP {\bf 1105}, 032 (2011)
  [arXiv:1103.1714 [hep-th]].

\bibitem{Kim:2013oba}
  K.~-Y.~Kim and M.~Taylor,
  ``Holographic d-wave superconductors,''
  arXiv:1304.6729 [hep-th].


\bibitem{MgB2-nature}
J. Nagamatsu, N. Nakagawa, T. Muranaka, Y.  Zenitani and J. Akimitsu,
"Superconductivity at 39 K in magnesium diboride". Nature 410 (6824): 63 (2001).

\bibitem{MgB2}
X.-X. Xi, ``Two-band superconductor magnesium diboride", Rep. Prog. Phys. 71 116501 (2008).

\bibitem{Berg:2009dga}
  E.~Berg, E.~Fradkin, S.~A.~Kivelson and J.~M.~Tranquada,
  ``Striped superconductors: how spin, charge and superconducting orders intertwine in the cuprates,''
  New J.\ Phys.\  {\bf 11}, 115004 (2009).


\bibitem{Zaanen:2010yk}
  J.~Zaanen,
  ``A Modern, but way too short history of the theory of superconductivity at a high temperature,''
  arXiv:1012.5461 [cond-mat.supr-con].

\bibitem{Fujimoto}
S.~Fujimoto,
"Electron Correlation and Pairing States in Superconductors without Inversion Symmetry,"
J.~Phys.~Soc.~Jpn.{\bf 76} , 051008 (2007),
[arXiv: [cond-mat.supr-con].

\bibitem{Goswami:2013wta}
  P.~Goswami and B.~Roy,
  ``Axionic superconductivity in three dimensional doped narrow gap semiconductors,''
  arXiv:1307.3240 [cond-mat.supr-con].


\bibitem{Basu:2010fa}
  P.~Basu, J.~He, A.~Mukherjee, M.~Rozali and H.~-H.~Shieh,
  ``Competing Holographic Orders,''
   JHEP {\bf 1010}, 092 (2010)  [arXiv:1007.3480 [hep-th]].

\bibitem{Cai:2013wma}
  R.~-G.~Cai, L.~Li, L.~-F.~Li and Y.~-Q.~Wang,
  ``Competition and Coexistence of Order Parameters in Holographic Multi-Band Superconductors,''  JHEP {\bf 1309}, 074 (2013)  [arXiv:1307.2768 [hep-th]].



\bibitem{Nie:2013sda}
  Z.~-Y.~Nie, R.~-G.~Cai, X.~Gao and H.~Zeng,
  ``Competition between the s-wave and p-wave superconductivity phases in a holographic model,''
  JHEP {\bf 1311}, 087 (2013)  [arXiv:1309.2204 [hep-th]].


\bibitem{Huang:2011ac}
  C.~-Y.~Huang, F.~-L.~Lin and D.~Maity,
  ``Holographic Multi-Band Superconductor,''
  Phys.\ Lett.\ B {\bf 703}, 633 (2011)  [arXiv:1102.0977 [hep-th]].

\bibitem{Krikun:2012yj}
  A.~Krikun, V.~P.~Kirilin and A.~V.~Sadofyev,
  ``Holographic model of the $S^{\pm}$ multiband superconductor,''
  JHEP {\bf 1307}, 136 (2013)  [arXiv:1210.6074 [hep-th]].


\bibitem{Musso:2013ija}
  D.~Musso,
  ``Competition/Enhancement of Two Probe Order Parameters in the Unbalanced Holographic Superconductor,''
   JHEP {\bf 1306}, 083 (2013)  [arXiv:1302.7205 [hep-th]].


\bibitem{Nitti:2013xaa}
  F.~Nitti, G.~Policastro and T.~Vanel,
  ``Dressing the Electron Star in a Holographic Superconductor,''
  JHEP {\bf 1310}, 019 (2013)
  [arXiv:1307.4558 [hep-th]].

\bibitem{Liu:2013yaa}
  Y.~Liu, K.~Schalm, Y.~-W.~Sun and J.~Zaanen,
  ``Bose-Fermi competition in holographic metals,''
  JHEP {\bf 1310}, 064 (2013)
  [arXiv:1307.4572 [hep-th]].


\bibitem{Amado:2013lia}
  I.~Amado, D.~Arean, A.~Jimenez-Alba, L.~Melgar and I.~Salazar Landea,
  ``Holographic s+p Superconductors,''
   Phys.\ Rev.\ D {\bf 89}, 026009 (2014)  [arXiv:1309.5086 [hep-th]].  


\bibitem{Donos:2012yu}
  A.~Donos, J.~P.~Gauntlett, J.~Sonner and B.~Withers,
  ``Competing orders in M-theory: superfluids, stripes and metamagnetism,''
   JHEP {\bf 1303}, 108 (2013)  [arXiv:1212.0871 [hep-th]].


\bibitem{Wen:2013ufa}
  W.~-Y.~Wen, M.~-S.~Wu and S.~-Y.~Wu,
  ``A Holographic Model of Two-Band Superconductor,''
   Phys.\ Rev.\ D {\bf 89}, 066005 (2014)  [arXiv:1309.0488 [hep-th]].


\bibitem{Amoretti:2013oia}
  A.~Amoretti, A.~Braggio, N.~Maggiore, N.~Magnoli and D.~Musso,
  ``Coexistence of two vector order parameters: a holographic model for ferromagnetic superconductivity,''
  JHEP {\bf 1401}, 054 (2014)
  [arXiv:1309.5093 [hep-th]].

\bibitem{Donos:2013woa}
  A.~Donos, J.~P.~Gauntlett and C.~Pantelidou,
  ``Competing p-wave orders,''
  Class.\ Quant.\ Grav.\  {\bf 31}, 055007 (2014)
  [arXiv:1310.5741 [hep-th]].


\bibitem{Nishida:2014lta}
  M.~Nishida,
  ``Phase Diagram of a Holographic Superconductor Model with s-wave and d-wave,''  arXiv:1403.6070 [hep-th].

\bibitem{Cai:2012nm}
  R.~-G.~Cai, S.~He, L.~Li and Y.~-L.~Zhang,
  ``Holographic Entanglement Entropy on P-wave Superconductor Phase Transition,''
   JHEP {\bf 1207}, 027 (2012)  [arXiv:1204.5962 [hep-th]].


\bibitem{Pallab:2014}
  P.~Goswami, B.~Roy,
  ``Axionic superconductivity in three dimensional doped narrow gap semiconductors,''
   Phys. Rev. B {\bf 90}, 041301(R)(2014)  	[arXiv:1307.3240 [cond-mat.supr-con]].

\end{thebibliography}
\end{document}